\documentclass[a4paper]{jpconf}
\usepackage[dvipdfm]{graphicx}
\usepackage{wrapfig}
\usepackage{amsmath,amssymb}

\newcommand{\bra}[1]{\langle {#1} |}
\newcommand{\ket}[1]{| {#1} \rangle}
\newcommand{\inproduct}[2]{\langle #1 | #2 \rangle}

\begin{document}
\title{Density functional approaches to atomic nuclei\footnote{
This work is supported by Grant-in-Aid for Scientific Research
in Japan (Nos. 21340073 and 20105003).
}}

\author{Takashi Nakatsukasa}

\address{RIKEN Nishina Center, Wako, 351-0198, Japan}

\ead{nakatsukasa@riken.jp}

\begin{abstract}
Nuclear mean-field models are briefly reviewed
to illustrate its foundation and necessity of state dependence in
effective interactions.
This state dependence is successfully taken into account by the density
dependence, leading to the energy density functional.
Recent results for photoabsorption cross sections in
spherical and deformed Nd isotopes are shown.
\end{abstract}

\section{Introduction}
\label{sec:intro}

The nucleus is a self-bound quantum system which presents a rich variety
of phenomena.
It is composed of fermions of spin $1/2$ and isospin $1/2$,
called nucleons (protons and neutrons),
interacting with each other through
a complex interaction with a short-range repulsive core \cite{BM69}.
Remarkable experimental progress in production and study of exotic nuclei
requires us to construct a theoretical model with higher accuracy and
reliability.
Extensive studies have been made in the past,
to introduce models and effective interactions to describe
a variety of nuclear phenomena and to understand basic nuclear dynamics
behind them \cite{BM69,RS80}.
Simultaneously, significant efforts have been made in the microscopic
foundation of those models.
For light nuclei,
the ``first-principles'' large-scale computation,
starting from the bare nucleon-nucleon (two-body \& three-body) forces,
is becoming a current trend in theoretical nuclear physics.

Although the ab-initio-type approaches
have recently shown a significant progress,
they are still limited to nuclei with the small mass number.
In contrast, the density functional model is a leading theory for describing
nuclear properties of heavy nuclei and perhaps the only theory capable of
describing all nuclei and nuclear matter with a single universal
energy density functional.
In the nuclear physics, it is often called the self-consistent mean-field
model, because of a historical development based on the
Brueckner-Hartree-Fock theory and introduction of the effective interaction.
I briefly review basic properties of nuclei and discuss whether those
properties can be understood by a simple independent-particle model.
Then, I recapitulate developments in the microscopic many-body theory
leading to the nuclear density
functional model with a Skyrme energy functional.
Here, the saturation property plays a key role to understand
the nuclear force and the effective interaction.

Intensive studies in nuclear density functional models in recent years
have produced numerous results and new insights into nuclear structure
\cite{BHR03,LPT03}.
However, it is impossible to review all of them in this short paper.
Thus, I will present a result of our recent study
with the time-dependent density functional approach,
on the photoabsorption cross sections in the rare-earth nuclei \cite{YN11}.

\section{Independent particle model for nuclei}
\label{sec:IPM}

Nuclei are known to be well characterized by the saturation property.
Namely, they have an approximately constant density
$\rho_0\approx 0.17$ fm$^{-3}$,
and a constant binding energy per particle $B/A\approx 16$ MeV.\footnote{
This is the extrapolated value for the infinite nuclear matter
without the surface and the Coulomb energy.
The observed values for finite nuclei are
$B/A\approx 8$ MeV.}
In this section, I show
that the nuclear saturation property has a great impact on
nuclear models.
Especially, it is inconsistent with the independent-particle model of nuclei
with a ``naive'' average (mean-field) potential.

There are many evidences for the fact that
the mean-free path of nucleons
is larger than the size of nucleus.
In fact, the mean free path depends on the nucleon's energy,
and becomes larger for lower energy \cite{BM69}.
Therefore, it is natural to assume that
the nucleus can be primarily approximated by the
independent-particle model with an average one-body potential.
The crudest approximation is the degenerate Fermi gas of the same number of
protons and neutrons ($Z=N=A/2$).
The observed saturation density of $\rho_0\approx 0.17$ fm$^{-3}$
gives the Fermi momentum, $k_F\approx 1.36$ fm$^{-1}$,
that leads to the Fermi energy (the maximum kinetic energy),
$T_F=k_F^2/2M\approx 40$ MeV.

First, I show that the independent-particle model
with a constant attractive potential $V<0$ cannot describe
the nuclear saturation property.
It follows from the simple arguments.
The constancy of $B/A$ means that it is approximately equal to the
separation energy of nucleons, $S$.
In the independent-particle model, it is estimated as
\begin{equation}
\label{B1}
S \approx B/A \approx -(T_F + V) .
\end{equation}
Since the binding energy is $B/A\approx 16$ MeV,
the potential $V$ is about $-55$ MeV.
It should be noted that the relatively small separation energy is
the consequence of the significant cancellation between
kinetic and potential energies.
The total (binding) energy is given by
\begin{equation}
\label{B2}
-B = \sum_{i=1}^A \left( T_i + \frac{V}{2} \right)
        = A \left( \frac{3}{5}T_F + \frac{V}{2} \right) ,
\end{equation}
where we assume that the average potential results from a two-body
interaction.
The two kinds of expressions for $B/A$, Eqs. (\ref{B1}) and (\ref{B2}),
lead to $T_F\approx -5V/4\approx 70$ MeV,
which is different from the previously estimated value ($\sim 40$ MeV).
Moreover, it contradicts the fact
that the nucleus is bound ($T_F < |V|$).

To reconcile
the independent-particle motion with the saturation property of the nucleus,
the nuclear average potential should be state dependent.
Allowing the potential $V_i$ depend on the state $i$,
the potential $V$ should be replaced by that for the highest occupied
orbital $V_F$ in Eq. (\ref{B1}),
and by its average value $\langle V \rangle$ in
the right-hand side of Eq. (\ref{B2}).
Then, we obtain the following relation:
\begin{equation}
\label{V_F}
V_F \approx \langle V \rangle + T_F/5 + B/A .
\end{equation}
Therefore, the potential $V_F$ is shallower 
than its average value.

Weisskopf suggested the momentum-dependent potential $V$, which can be
expressed in terms of an effective mass $m^*$ \cite{Wei57}:
\begin{equation}
\label{mom_dep_pot}
V_i=U_0+U_1\frac{k_i^2}{k_F^2} .
\end{equation}
Actually, if the mean-field potential is non-local, it can be
expressed by the momentum dependence.
Equation (\ref{mom_dep_pot})
leads to the effective mass, $m^*/m = (1+U_1/T_F)^{-1}$.
Using Eqs. (\ref{B1}), (\ref{V_F}), and (\ref{mom_dep_pot}),
we obtain the effective mass as
\begin{equation}
\label{m*/m}
\frac{m^*}{m} = \left\{ \frac{3}{2} + \frac{5}{2}\frac{B}{A}\frac{1}{T_F}
 \right\}^{-1} \approx 0.4 .
\end{equation}
Quantitatively, this value disagrees with the experimental data.
The empirical values of the effective mass
vary according to the energy of nucleons,
$0.7 \lesssim m^*/m \lesssim 1$,
however, they are almost twice larger than
the value in Eq. (\ref{m*/m}).
As far as we use a normal two-body interaction,
this discrepancy should be present in the mean-field calculation
with any interaction,
because Eq. (\ref{m*/m}) is valid in general
for a saturated self-bound system.
Therefore, the conventional models cannot simultaneously
reproduce the most basic properties of nuclei; the binding energy and
the single-particle property.
This suggest the importance of the state-dependent
effective interaction, which will be discussed in Sec. \ref{sec:DDHF}.

\section{Nucleon-nucleon interaction (nuclear force)}
\label{sec:NN}

The saturation property of nuclear density indicates the balance
between attractive and repulsive contributions to nuclear binding
energy.
One source of such repulsive effects is the nucleonic kinetic energy
of the Fermi gas.
However, its contribution per particle is proportional to $\rho^{2/3}$,
which is not strong enough to resist against the collapse caused by
the attractive force between nucleons.
Therefore, the nucleonic interaction must contain a repulsive element.
Indeed, the phase-shift analysis on the nucleon-nucleon scattering
at high energy ($E>250$ MeV) reveals a short-range strong repulsive
core in the nucleonic force.
The radius of the repulsive core is approximately $c\approx 0.5$ fm.
This strong repulsive core prevents the nucleons approaching
closer than the distance $c$, which produces a strong two-body correlation,
$\rho^{(2)}(\vec{r}_1,\vec{r}_2)\approx 0$ for $|\vec{r}_1-\vec{r}_2|<c$.
The attractive part of the interaction has a longer range, which can
be characterized by the pion's Compton wave length $\lambda_\pi$,
and is significantly weaker than the repulsion.
Thus, a naive application of the mean-field calculation fails to bind
the nucleus, since the mean-field approximation cannot take account
of such strong two-body correlations.

At first sight,
this seems inconsistent with the experimental observations.
As I mentioned in Sec.~\ref{sec:IPM},
there are many experimental evidences for the independent-particle
motion in nuclei.
We may intuitively understand that it is due to the fact that the
nucleonic density is significantly smaller than $1/c^3$.
Therefore, the collisions by the repulsive core rarely occur and
the system can be approximately described in terms of the
independent-particle motion.
Furthermore, the effects of the Pauli principle hinder the collisions,
since the nucleons cannot be scattered into occupied states.
Although the repulsive-core collisions are experienced by only
a small fraction of nucleons ($\sim \rho_0 c^3$),
each collision carries a large amount of energy.
Therefore, the repulsive core provides an important contribution to the
total energy and are responsible for the saturation.

Another important factor for the independent particle motion
is the strong quantum nature due to the weakness of the attractive
part of the nuclear force.
The importance of the quantum nature can be measured by the
magnitude of the zero-point kinetic energy compared to that of the
interaction.
If the attractive part of the nuclear force were much stronger
than the unit of $\hbar^2/Mc^2$,
the quantum effect would disappear and
each nucleon would stay at the bottom of the interaction potential
(cf. Fig. 2-36 in Ref. \cite{BM69}).
Then, the nucleus would crystallize at low temperature.
In reality, the attraction of the nuclear force is so weak
that it barely produces many-nucleon bound states at the relatively
low density.
In Sec. \ref{sec:IPM}, I have shown that, in nuclear binding energy,
there is a strong cancellation
between the positive kinetic energy and the negative potential energy.
The nucleonic kinetic energy plays an important role in many phenomena
in nuclei,
which can be described in terms of the
independent-particle motion.

\section{Density-dependent Hartree-Fock method and energy density functional}
\label{sec:DDHF}

The nuclear matter theory pioneered by Brueckner
gives a hint for a solution for the inconsistency between
the nuclear saturation and the independent-particle model.
Details of the theory can be found at Refs. \cite{Day67,RS80}.
The independent-particle motion under the presence of the interaction
with a repulsive core was qualitatively discussed in Sec. \ref{sec:NN}.
The Brueckner theory may provide a first step toward the quantitative
treatment to understand the saturation property and the
independent-particle motion in nuclei.

The basic ingredient of the Brueckner theory is a two-body scattering
matrix of particle 1 and 2 inside nucleus caused by the nuclear force $v$,
\begin{equation}
\label{G-matrix}
G(\omega)\equiv v+v \frac{Q}{\omega-Q(T_1+T_2)Q} G(\omega),
\end{equation}
where $T_i$ is the kinetic energy of particle $i$,
$Q$ is the Pauli-exclusion operator to restrict the intermediate states,
and $\omega$ is called a starting energy that depends on energies of
particle 1 and 2.
This is called $G$-matrix \cite{BG57}.
The $G$-matrix renormalizes high-momentum components in the bare nuclear
force and becomes an effective interaction in nuclei under
the independent-pair approximation.
The $G$-matrix reflects an underlying structure
of the independent many-nucleon system through the operator $Q$ and the
starting energy $\omega$.
Inevitably, the $G$-matrix becomes state (structure) dependent.

Since the short-range singularity is renormalized in the $G$-matrix,
we can calculate the total energy in the independent-particle
(mean-field) model, analogous to Eq. (\ref{B2}).
\begin{equation}
-B=\sum_{i=1}^A \left\{ T_i +
 \frac{1}{2} \sum_{j=1}^A
\bar{G}_{ij,ij}(\omega_{ij})
\right\}
\end{equation}
where $\omega_{ij}=\epsilon_i+\epsilon_j$,
defines the self-consistency condition for the Brueckner's
single-particle energies,
and $\bar{G}_{ij,ij}\equiv G_{ij,ij}-G_{ij,ji}$.
This is called Brueckner-Hartree-Fock (BHF) theory.
The validity of the BHF theory is measured by the wound integral
$\kappa=\inproduct{\psi-\phi}{\psi-\phi}$, where $\ket{\phi}$ is
an unperturbed two-particle wave function and $\ket{\psi}$ is
a correlated two-particle wave function in nucleus.
$\kappa$ is known to be of the order of 15 \%.
The BHF calculation was successful to describe the nuclear saturation,
however, could not reproduce simultaneously $B/A$ and $\rho_0$,
known as a problem of the Coester band \cite{Day81a}.
Its applications to finite nuclei also quantitatively failed to reproduce
the energy, radius, and density in the ground state.

These problems are somewhat miraculously solved by the
density-dependent Hartree-Fock (DDHF) theory by Negele \cite{Neg70}.
Starting from a realistic $G$-matrix, first, the local density
approximation is introduced, using the expressions for
the Pauli operator
\begin{equation}
\bra{\vec{r}_1\vec{r}_2}Q\ket{\vec{r}'_1\vec{r}'_2}
= \left\{ \delta(\vec{r}_1-\vec{r}'_1) - \rho(\vec{r}_1-\vec{r}'_1) \right\}
\left\{ \delta(\vec{r}_2-\vec{r}'_2) - \rho(\vec{r}_2-\vec{r}'_2) \right\} ,
\end{equation}
and the average single-particle energy $\epsilon[\rho(\vec{r})]$.
Then, a short-range part of the $G$-matrix, which is not fully understood,
is phenomenologically added to the
energy expression to quantitatively fit the saturation property,
and finally, the total energy is treated variationally.
This procedure is called the density matrix expansion (DME) \cite{NV72}.
The state dependence of the $G$-matrix is now replaced by the
density dependence.
The final result for the energy is of the form
\begin{equation}
E[\rho] = \int d\vec{R} H(\vec{R}) ,
 \quad H(\vec{R})=H[\rho(\vec{R})]=H[\psi^*,\psi] ,
\end{equation}
which is completely analogous to the Hamiltonian density of the
Skyrme energy functional \cite{VB72}.

The essential aspect of the DDHF comes from the density dependence
and the variational treatment.
The variation of the total energy with respect to the density
contains re-arrangement potential, $\partial V_{\rm eff}[\rho]/\partial\rho$,
which appear due to the density dependence of the effective force
$V_{\rm eff}[\rho]$.
These terms turn out to be crucial to obtain the saturation condition.
Now, the expression for the total energy, Eq. (\ref{B2}), should be modified
to include the re-arrangement effect.
This resolves the previous issue, then provides a consistent
independent-particle description for the nuclear saturation.

In summary, the failure in the mean-field description of nuclei
using phenomenological effective interactions can be traced back
to the missing state (structure) dependence.
The DDHF takes into account the state dependence
in terms of the density dependence.
The energy functional obtained by the DME is essentially identical
to the Skyrme energy functional.
This provides a foundation for the nuclear energy functional.

\section{Applications to giant resonances}
\label{sec:GR}

The giant resonance is a typical collective motion in nuclei, which
exhausts a major part of the sum-rule value.
They are also related to the basic properties of nuclear matter, such as
incompressibility $K_\infty$, effective mass $m^*$, etc.
For instance, the nuclear matter incompressibility is extracted from
properties of the giant monopole resonances \cite{Bla80},
\begin{equation}
K_\infty = 210 \pm 30 \mbox{ MeV} .
\end{equation}
This gives a restriction on the density dependence of
the phenomenological short-range repulsive part of the energy functional.
The effective mass deduced from the analysis on the giant quadrupole
resonances is \cite{Bla80}
\begin{equation}
m^*/m\approx 0.8\sim 1 .
\end{equation}
These values are consistent with the DDHF in Sec. \ref{sec:DDHF},
but inconsistent with the naive mean-field value of Eq. (\ref{m*/m}).

\begin{wrapfigure}{r}{6cm}
\includegraphics[width=6cm]{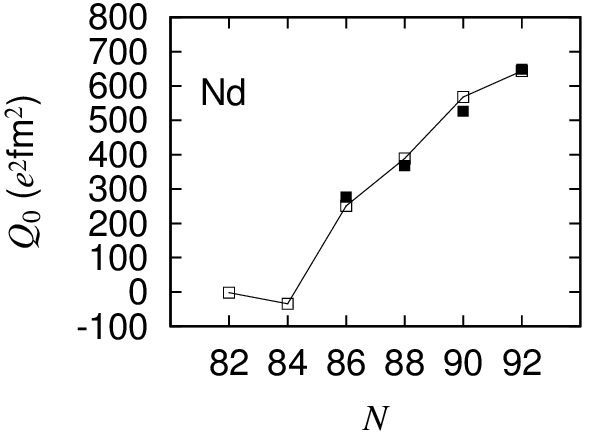}
\caption{Calculated (open squares) and experimental (filled)
intrinsic quadrupole moment for Nd isotopes \cite{YN11}.}
\label{fig:Nd_def}
\end{wrapfigure}
The giant resonances can be reasonably described by a small-amplitude
approximation of the time-dependent version of the DDHF.
In this approach, again, the re-arrangement terms, such as
$\partial V_{\rm eff}[\rho]/\partial\rho$ and
$\partial^2 V_{\rm eff}[\rho]/\partial\rho^2$ should be
consistently taken into account.
The theory can be regarded as
the time-dependent density-functional
theory, founded by Runge and Gross \cite{RG84}.

A modern energy functional for nuclei is a functional of many kinds
of density, such as kinetic $\tau(\vec{r})$
and spin-orbit density $\vec{J}(\vec{r})$.
In addition, to describe superfluid nuclei with pairing correlation,
we need to add the pair (abnormal) density $\kappa(\vec{r})$.
These densities are collectively denoted as $\tilde{\rho}$ in the
followings.
Variation of the total energy, $E[\tilde{\rho}]$,
leads to the Hartree-Fock-Bogoliubov equation:
\begin{equation}
\label{HFB}
H[\Psi,\Psi^*] \ket{\Psi_\mu} =
E_\mu \ket{\Psi_\mu} ,
\end{equation}
where $E_\mu$ and
$\ket{\Psi_\mu}$ are quasi-particle energies
and states, respectively \cite{RS80}.
$\ket{\Psi_\mu}$ is composed of two components;
the upper $\ket{U_\mu}$ and
the lower one $\ket{V_\mu}$.
The solution of Eq. (\ref{HFB}) defines the normal density
$\rho(\vec{r})=\sum_\mu V_\mu(\vec{r}) V_\mu^*(\vec{r})$,
the pair density
$\kappa(\vec{r})=\sum_\mu U_\mu(\vec{r}) V_\mu(\vec{r})$,
and other densities at the ground state.
Since $h[\Psi,\Psi^*]$ depends on these densities,
Eq. (\ref{HFB}) must be solved in a self-consistent way.
Minimization of the energy density functional may lead to
a spontaneous breaking of symmetry.
An example is given in Fig. \ref{fig:Nd_def} for Nd isotopes \cite{YN11}.
The intrinsic quadrupole moment calculated with the Skyrme
functional of SkM* is compared with the
experimental data.
At $N=82$, the nucleus at the ground state is spherical $Q_0=0$,
while for $N=86\sim 92$,
the deformation gradually develops.
The observed ground-state deformations deduced from the transition probability
$B(E2; 2^+ \rightarrow 0^+)$ are nicely reproduced.
Note that there are no adjustable parameters in this calculation.

\begin{wrapfigure}{l}{5cm}
\includegraphics[width=5cm]{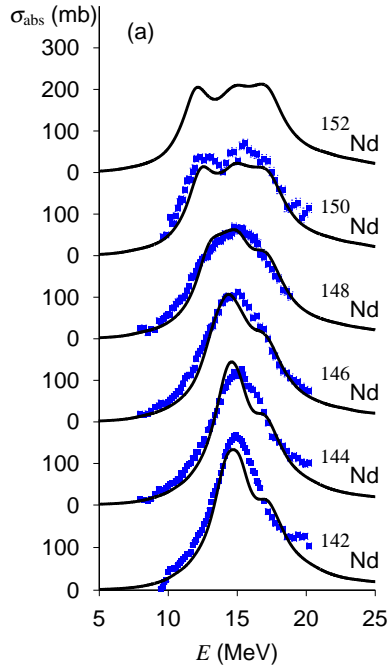}
\caption{Calculated (lines) and experimental (symbols) photoabsorption cross
sections for Nd isotopes \cite{YN11}.}
\label{fig:Nd}
\vspace{-0.5cm}
\end{wrapfigure}
For a study of the giant resonances,
we need to extend the energy functional
to include the time-odd densities, such as the spin density $\vec{s}(\vec{r})$
and the current density $\vec{j}(\vec{r})$.
Now, these densities are time dependent.
The time-dependent version of Eq. (\ref{HFB}) is
\begin{equation}
\label{TDHFB}
i\frac{\partial}{\partial t} \ket{\Psi_\mu(t)} =
H[\Psi(t), \Psi^*(t)] \ket{\Psi_\mu(t)} .
\end{equation}
Assuming the oscillation with a fixed frequency,
this equation is linearized with respect to the fluctuation of
the densities around those at the ground state.
This leads to the matrix form of the equation identical to the
quasi-particle random-phase approximation (QRPA) \cite{RS80}.
\begin{equation}
\label{QRPA}
\sum_{\gamma\delta}
\begin{pmatrix}
A_{\alpha\beta,\gamma\delta} & B_{\alpha\beta,\gamma\delta} \\
-B_{\alpha\beta,\gamma\delta} & -A_{\alpha\beta,\gamma\delta}
\end{pmatrix}
\begin{pmatrix}
X_{\gamma\delta} \\
Y_{\gamma\delta}
\end{pmatrix}
= \hbar\omega
\begin{pmatrix}
X_{\alpha\beta} \\
Y_{\alpha\beta}
\end{pmatrix} .
\end{equation}
The QRPA matrix, $A$ and $B$, are calculated in the quasi-particle
basis, then the normal modes of excitation and their energies 
are obtained from Eq. (\ref{QRPA}).

The calculated photoabsorption cross sections for Nd isotopes
are shown in Fig.~\ref{fig:Nd}.
For spherical nuclei ($N=82$ and 84), the photoabsorption has a single
peak for the photon energy of $E\approx 15$ MeV.
The increase of the neutron number results in the
broadening of the peak, which well agree with the experimental data.
This is due to the increase of the ground-state deformation shown in
Fig. \ref{fig:Nd_def}.
The calculated energy-weighted sum-rule values for these isotopes are
about 40 \% larger than the classical Thomas-Reiche-Kuhn sum-rule value,
because of the momentum and isospin dependence of the mean field
(Kohn-Sham) potential.

\section{Summary}

The nuclear density functional approaches were developed as the mean-field
theory with the density-dependent effective interactions.
The density-dependent Hartree-Fock theory succeeded to
describe both the total energy and single-particle properties, simultaneously.
In contrast,
the normal mean-field calculation with a phenomenological two-body
interaction fails.
The density-dependence is a key ingredient to understand
the nuclear saturation and the independent-particle motion.
Numerical applications were shown for giant resonances
in shape transitional nuclei.

\section*{References}
\bibliographystyle{iopart-num}
\bibliography{nuclear_physics,chemical_physics,myself,current}

\end{document}